\documentclass{jetpl}
\usepackage[T2A]{fontenc}
\usepackage[english]{babel}
\usepackage{graphicx}
\usepackage{epsfig}
\twocolumn

\lat

\title{Observation of the magnetic domain structures in Cu$_{0,47}$Ni$_{0,53}$ thin films at low
temperatures}

\rtitle{Observation of the magnetic domain structures\ldots}

\sodtitle{Observation of the magnetic domain structures in
Cu$_{0,47}$Ni$_{0,53}$ thin films at low temperatures}

\author{I.\,S.\,Veshchunov, V.\,A.\,Oboznov, A.\,N.\,Rossolenko,
A.\,S.\,Prokofiev, L.\,Ya.\,Vinnikov \thanks{e-mail:
vinnik@issp.ac.ru}, A.\, Yu.\, Rusanov and D.\,V.\,Matveev}

\rauthor{I.\,S.\,Veshchunov, V.\,A.\,Oboznov, A.\,N.\,Rossolenko,
A.\,S.\,Prokofiev, L.\,Ya.\,Vinnikov, Yu.\, Rusanov and
D.\,V.\,Matveev}

\sodauthor{Veshchunov, Oboznov, Rossolenko, Prokofiev, Vinnikov,
Rusanov and Matveev}

\address{Institute of Solid State Physics RAS, 142432 Chernogolovka, Moscow distr., Russia}

\dates{22 October 2008}{*}

\abstract{We report on the first experimental visualization of
domain structure in films of weakly ferromagnetic
Cu$_{0,47}$Ni$_{0,53}$ alloy with different thickness at liquid
helium temperatures. Improved high-resolution Bitter decoration
technique was used to map the magnetic contrast on the top of the
films well below the Curie temperature T$_{Curie}$ ($\sim$ 60 K). In
contrast to magnetic force microscopy, this technique allowed
visualization of the domain structure without its disturbance while
the larger areas of the sample were probed. Maze-like domain
patterns, typical for perpendicular magnetic anisotropy, were
observed. The average domain width was found to be about 100 nm.}


\begin{document}

\maketitle

The interplay between superconductivity and ferromagnetism leads to
a number of interesting phenomena~\cite{1,2}, which can be utilized
in various applications. The interest of investigating the domain
structure of weakly ferromagnetic Cu$_{1 - x}$Ni$_{x}$ (x$\sim $0.5)
alloys, in particular, is caused by their active use in thin film
superconductor (S)/ ferromagnet (F) heterostructures. The most
promising microelectronic devices based on such heterostructures are
basic elements of digital rapid single flux quantum (RSFQ) and
quantum logic circuits. Apart from that, using Cu$_{1 - x}$Ni$_{x}$
films as weak ferromagnets in fundamental S/F system properties
research looks quite promising, which was confirmed in numerous
theoretical and experimental investigations~\cite{2,21,22,23,3,3a}.
The physical properties of CuNi alloys are relatively well studied.
The average magnetic moment and the Curie temperature of uniform
CuNi alloys decrease linearly with Ni concentration and both
approach zero at $\sim $45 at.{\%} Ni content. The magnetism of CuNi
films is weaker than that in bulk material. Although Cu$_{1 - x
}$Ni$_{x}$ films with Ni concentration close to the critical value
seem to be good candidates for using them in S/F proximity systems,
at the same time they have several disadvantages. One of them is
that the film structure is very sensitive to the fabrication
conditions. In particular, the homogeneity of the sputtered films is
not ideal, (at least close to x=0.5) since there is a tendency of
Ni-rich clusters forming~\cite{6}. But the features of CuNi films
domain structure, which have huge influence on the transport and
magnetic properties of S/F systems~\cite{3} have never been
revealed. In this paper we present results of studying magnetic
domains in thin films of Cu$_{0,47}$Ni$_{0,53}$ (hereafter called
CuNi) alloy.

In the past decade magnetic force microscopy (MFM) has become a
well-established technique for the observation of the distribution
of magnetic domains with submicron
resolution~\cite{31,32,322,323,33,34}. It is widely used, except for
the cases when MFM magnetic probe might bring distortions into the
scanned image. That can happen, for instance, because of the sample
local magnetization change by the probe itself during the scan. It's
known that even well below the Curie temperature, the coercive field for
thin films of some ferromagnets (CuNi in particular) might become almost zero,
so using even magnetically soft MFM probes can disturb the picture
of local magnetization when performing scans. Therefore, to study
domain structure in such films we used the improved low temperature
Bitter decoration technique~\cite{4}, which also has such additional
advantages as high spatial resolution and magnetic sensitivity. The
Bitter technique is based on the deposition of fine dispersed
magnetic nanoparticles, driven by the stray field gradients in the
vicinity of magnetic material, at places where the magnetic field is
higher. Decoration patterns can be examined by means of scanning
electron microscopy (SEM). The patterns provide no information about
the magnitude of the magnetization, yet in materials even with low
stray fields, Bitter patterns can quickly yield information about
the size and shape of domains of various types that might be
present.

\begin{figure}[!t]
\begin{center}
\centerline{\psfig{figure=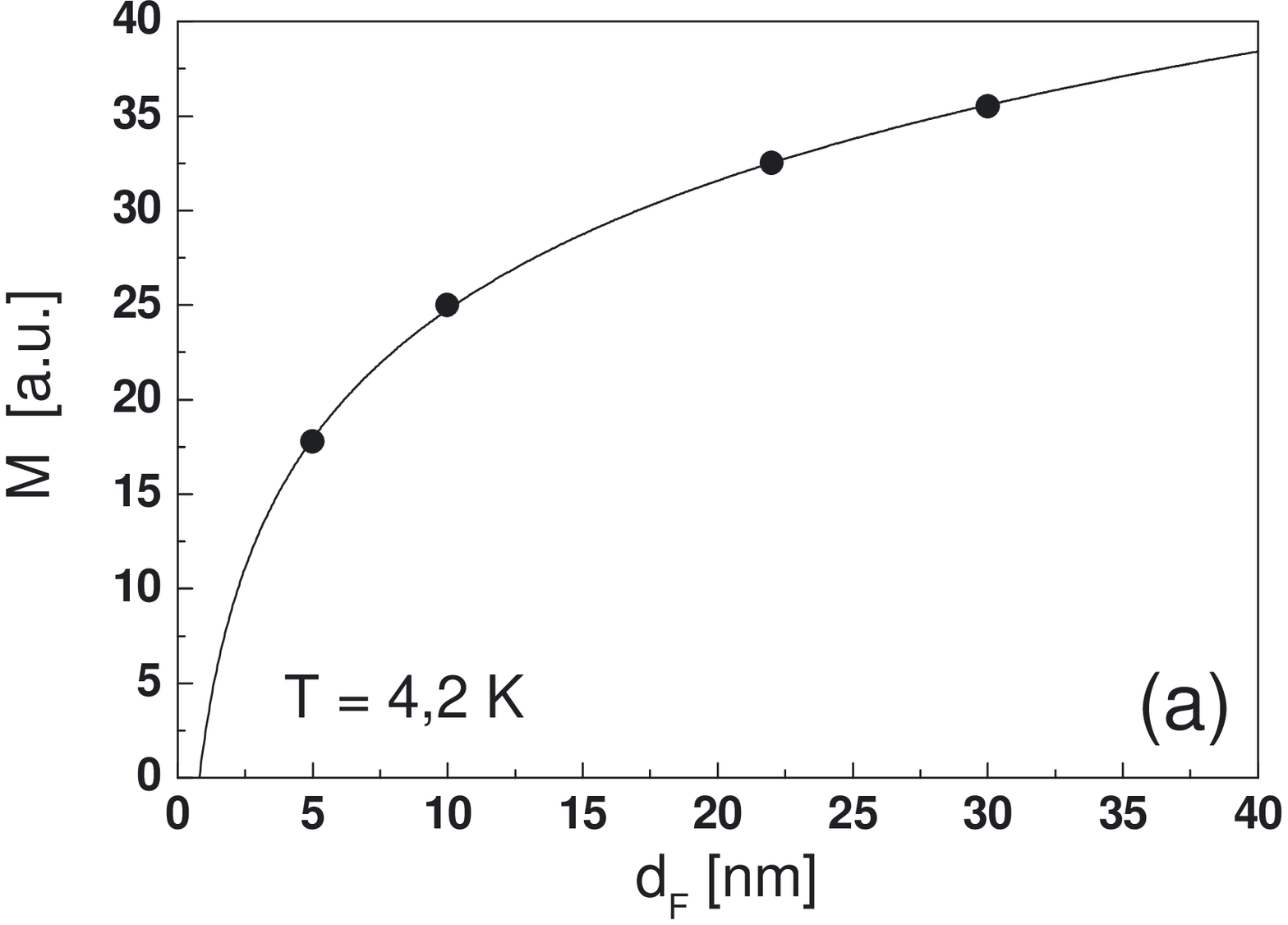, height =5cm}}
\centerline{\psfig{figure=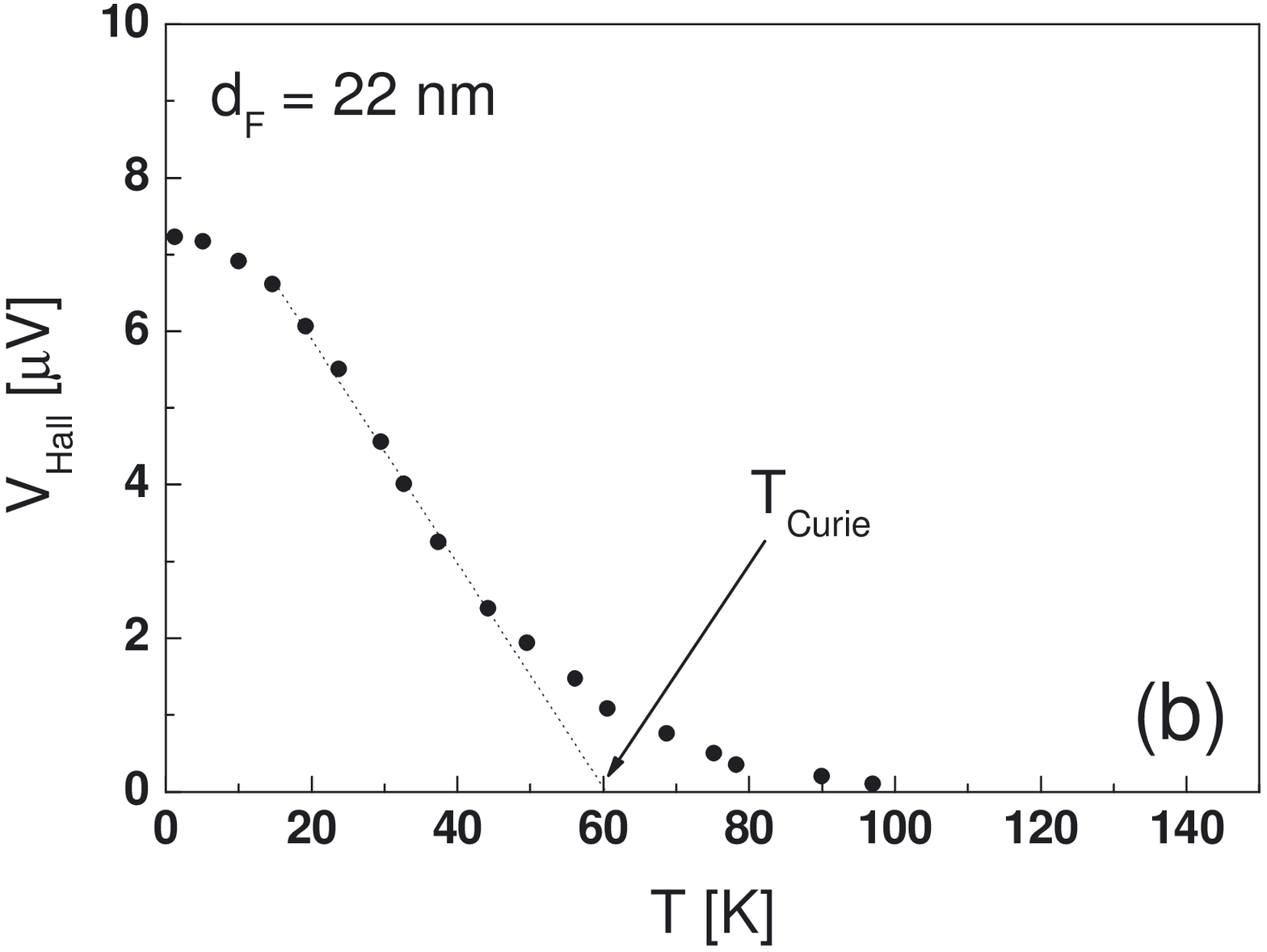, height =5cm}}
\bigskip
\caption{(a). Saturation magnetization $M$ as function of the film
thickness $d_F$ for Cu$_{0.47}$Ni$_{0.53}$. The dependence has a
logarithmic like behavior. The line between experimental points
serves to guide an eye. (b). Anomalous Hall voltage V$_{Hall}$
dependence with temperature $T$ for Cu$_{0.47}$Ni$_{0.53}$ film with
$d_F$=22 nm.} \label{fig1}
\end{center}
\end{figure}

For our experiments Cu$_{0,47 }$Ni$_{0,53}$ thin films were grown by
RF-sputtering in Ar atmosphere of P$_{Ar}$= 4$\times$10$^{ - 2
}$mbar on silicon substrates at room temperature. The deposition
rate was 0.25 nm/s. The Cu and Ni contents in the sputtered films
were determined by Rutherford Backscattering (RBS) analysis. It
confirmed that the Ni concentration in sputtered films was of the
same value as in used CuNi targets.

First, the magnetic properties of Cu$_{0.47}$Ni$_{0.53}$ films
structured in the shape of narrow bridges with thickness $d_F$
ranging from 5 to 30 nm were studied by measuring anomalous Hall
voltage V$_{Hall}$, which is proportional to the film magnetization
~\cite{7}. Fig.1a shows the dependence of V$_{Hall}$ corresponding
to the saturation magnetization of the sample with a particular
thickness on the sample thickness. In all cases the applied magnetic
field was perpendicular to the film surface. Typical temperature
dependence of the Hall voltage for the CuNi film with d$_F$=22 nm is
presented in Fig.1b. For all film thicknesses it appeared to be non
linear, with a weakly pronounced saturation at low temperatures and
tail-like behavior close to the T$_{Curie}$. T$_{Curie}$ for
different samples was estimated by extrapolating the linear part of
the V$_{Hall}(T)$ dependence as presented in Fig.1b. In order to
estimate the field range for the most effective magnetic domain
decoration the hysteresis magnetization loops were measured as well.
Results of magnetization reversal for Cu$_{0.47}$Ni$_{0.53 }$ film
with d$_{F}$=20 nm are presented in Fig.2. The measurements were
performed at 4.2K well below the T$_{Curie}$ (~60 K) of the sample.
Several field sweeps were performed with different values of maximum
field in the range of 150 - 700 Oe as shown in Fig.2. The coercive
field $H_{Coer}$ for those sweeps was found in the range between 50
- 150 Oe.

\begin{figure}[!t]
\begin{center}
\centerline{\psfig{figure=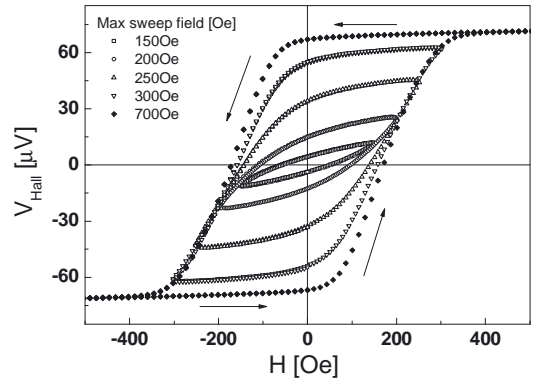, height =5cm}}
\bigskip
\caption{Hall voltage V$_{Hall}$ as function of applied magnetic field $H$ for Cu$_{0.47}$Ni$_{0.53}$ film with $d_F$=20nm measured at 4.2 K. Different curves correspond to different maximal sweep fields $H$ as indicated. For all curves $H$ was perpendicular to the film surface. Arrows show the direction of magnetic field sweep.} \label{fig2}
\end{center}
\end{figure}

Special consideration should be given to the CuNi thin film
decoration procedure. The sequence of the entire experiment can be
described as follows. The sample was initially cooled in zero
magnetic field down to 4,2 K. During the decoration the temperature
of the sample increased (of about 3-4 K) up to the decoration
temperature T$_{d}$ (the temperature measured by the resistive
thermometer in the end of the iron evaporation process). The first
series of decoration procedures was performed at magnetic fields
$H_{dec}$=100, 250, 300 Oe, on the virgin curve of the hysteresis
loop (see Fig.2). For each new field value a new sample
(geometrically identical to the previous one) was used. Additional
experiment was done to make sure that multiple domain situation
occurs after the magnetization switch. For that the magnetic field
was gradually increased from 0 to 300 Oe and then swept back to
$H_{dec}$= -150 Oe, which corresponds to $-H_{Coer}$ for this value
of maximum sweeping field, as shown in Fig.2. In all cases the
applied magnetic field was perpendicular to the sample surface.

Fig.3a, b, c, d present the distribution of the iron particles
mapping magnetic contrast related to the domain structure on the
surface of the sample obtained with SEM. Fig.3a, b, c show the
domain structure for decoration fields $H_{dec}$=100, 250 and 300 Oe
respectively. That, as it was mention before, corresponds to the
evolution of the domain state on the virgin curve of the hysteresis
loop. At the lowest applied field $H_{dec}$=100 Oe the decorated
domain structure implies practically demagnetized state on the
surface of the sample, which is believed to be perpendicular to the
spontaneous magnetization axis. Domains form a maze-like pattern
with a typical domain width of about 100 nm. Increasing the
decorating field $H_{dec}$ up to 250 Oe as indicated in Fig.3b
results in widening of the positive (magnetization is pointed up)
domains. Degradation of the pattern quality occurs because of the
decrease of the local field gradients at the film surface. The
domain structure (i.e. decorated magnetic contrast) almost
disappears when approaching $H_{dec}$=300 Oe, see Fig.3c. A
maze-like domain structure shows up clearly again, as it can be seen
from Fig.3d after magnetization switching, at the
$H_{dec}=H_{Coer}$= - 150 Oe.

\begin{figure}[!t]
\begin{center}
\centerline{\psfig{figure=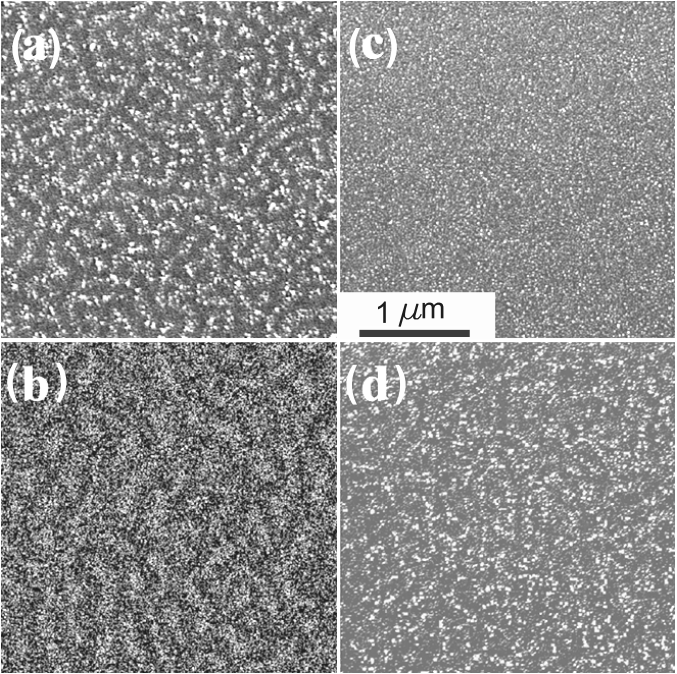, height =9cm}}
\bigskip
\caption{Evolution of the domain structure with external magnetic
field applied perpendicular to the film plane: (a) H=150 Oe, (b)
H=250 Oe, (c) H=300 Oe, (d) H= -150 Oe.} \label{fig3}
\end{center}
\end{figure}

The results of investigations reveal several advantages of using
CuNi alloys, which justify effectiveness of their utilization in
Josephson SFS junctions. First, for this particular ferromagnetic
material the exchange energy is relatively small (E$_{ex}$/k$_B$
$\sim$ 800 K and T$_{Curie}$ $\sim$ 60 K correspondingly). That
implies the superconducting order parameter decay length $\xi_{F1}$
and the period of its spatial oscillation $\xi_{F2}$ can be of the
order of several nanometers instead of $\sim$ 1 nm for the weak link
of Josephson SFS junction made of non-diluted ferromagnet.
Therefore, making Josephson SFS junctions with comparatively thick
F-layers using simple thin film technologies becomes possible.
Second, the domain structure of CuNi films has a spatial period of
about 0.1 $\mu$m, as the decoration experiments showed, which
provides a good averaging of the net magnetization in F-layer. This,
in its turn, allows to fabricate submicron ($\sim $0.2-0.3 $\mu$m)
SFS sandwiches without having the undesired macroscopic stray
fields. The final remark should be made about the possibility to
manipulate the Josephson characteristics of SFS junctions (critical
current, phase difference) with external magnetic field up to 20 Oe
without disturbing the domain structure of F-layer, since it has a
strong perpendicular anisotropy.

Several notes should be made on the decoration technique. Clearly,
in order to obtain the highest possible resolution of the method a
special care has to be taken of the decorating particle size as well
as of their magnetic properties \cite{16}. It's important to remind
that when the applied field is higher than the local stray fields
pointing up from the film surface, the magnetic particles become
polarized along the applied field before landing on the ferromagnet
surface~\cite{8}.

\begin{figure}[!t]
\begin{center}
\centerline{\psfig{figure=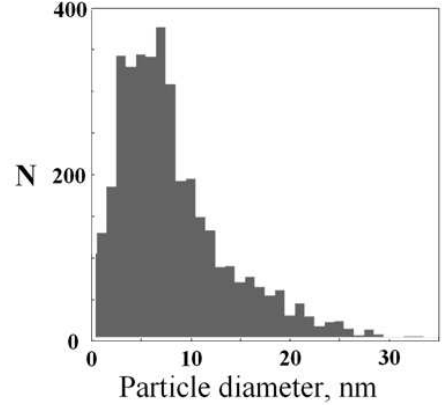, height =5.5cm}}
\bigskip
\caption{Size distribution of Fe decoration particles.
  N-number of Fe particles on the scanned area of $2 \mu m^2$. He gas
pressure was $\sim $2x10$^{ - 2}$ Torr.} \label{fig4}
\end{center}
\end{figure}

Therefore, the particles aggregate to the positive domain areas in
which the magnetization is aligned in the direction of an external
magnetic field while negative domain areas are free from iron
particles. Our additional experiments demonstrated that actual
particle size distribution strongly depends on the buffer He
pressure as was determined by SEM. Typical size distribution of the
particle size is presented in Fig.4. The highest resolution in the
range of 10-100nm was reached when the average particle size was of
about 10 nm, which is reached at (2-3)$\times$10$^{ - 2}$ Torr of
buffer He. The serious limitation of the decoration method lies in
fact that magnetic properties of the decorating particles strongly
depend on their size: the saturation magnetization decreases with
particle size (about 25{\%} of that in bulk iron for 10 nm size
particles~\cite{5}). Increasing the particle size implies larger
magnetic moment and thus higher magnetic sensitivity, but the
spatial resolution of the pattern gets reduced. Those particles with
larger magnetic moment tend to form irregular clusters on the sample
surface due to large interparticle interaction.

Recently, the perpendicular magnetic anisotropy in dc-magnetron
sputtered Ni$_{60}$Cu$_{40}$/Cu multilayers was detected by
hysteresis loop measurements for CuNi layer thickness between 4,2 nm
and 34 nm~\cite{9}. However, in that experiments the features of the
domain structure of CuNi films were not revealed.

In summary, the improved Bitter technique allowed visualizing the
domain structure of weakly ferromagnetic Cu$_{0,47}$Ni$_{0,53}$ on
large area (tens of square millimeters) at low temperatures. The
image of magnetic contrast on the top of Cu$_{0,47}$Ni$_{0,53}$
films was seen for the first time. It was experimentally shown that
thin CuNi films tend to have small scale domain structure. The films
with thickness in the range of 10-30 nm have perpendicular magnetic
anisotropy which results in maze-like domain patterns and nearly
rectangular hysteresis loop. The characteristic domain structure
scale is found to be about 100 nm. We are grateful to
V.\,V.\,Ryazanov and L.\, S.\, Uspenskaya for helpful discussions
and L.\, G.\, Isaeva for help in preparation of evaporators. This
work is supported by RFBR (07-02-00174), Joint Russian-Israeli
project MOST-RFBR (06-02-72025) and InQubit, {\bf Ltd.}


\begin{thebibliography}{99}
\bibitem{16} In our experiments done at low temperature magnetic particles were formed by evaporation of the
host material (in our case it was iron) from the surface of a
tungsten wire.
\end{thebibliography}
\end{document}